# Urban Spatial Structure and the Potential for Vehicle Miles Traveled Reduction: The Effects of Accessibility to Jobs within and beyond Employment Sub-centers

Marlon G. Boarnet[a], Xize Wang[b,*]

**Abstract**

This research examines the relationship between urban polycentric spatial structure and driving. We identified 46 employment sub-centers in the Los Angeles Combined Statistical Area and calculated access to jobs that are within and beyond these sub-centers. To address potential endogeneity problems, we use access to historically important places and transportation infrastructure in the early 20[th] century as instrumental variables for job accessibility indices. Our Two-stage Tobit models show that access to jobs is negatively associated with household vehicle miles traveled in this region. Among various accessibility measures, access to jobs outside sub-centers has the largest elasticity (-0.155). We examine the location of places in the top quintile of access to non-centered jobs and find that those locations are often inner ring suburban developments, near the core of the urban area and not far from sub-centers, suggesting that strategies of infill development that fill in the gaps between sub-centers, rather than focusing on already accessible downtowns and large sub-centers, may be the best land use approach to reduce VMT.

a. Price School of Public Policy, University of Southern California, Email: boarnet@usc.edu. OCRID: 0000-0002-0890-347X.
b. Institute of Urban and Regional Development, University of California, Berkeley, Email: wangxize316@gmail.com. OCRID: 0000-0002-4861-6002.
*. Corresponding author.

# 1. Introduction

This study investigates the relationship between polycentric urban structure and household-level travel behavior. The decentralization of population and employment has made many once-monocentric metropolitan areas more polycentric (Giuliano et al. 2007). There is a large literature studying the relationship between urban form and travel behavior. However, the effect of a polycentric urban structure on passenger travel has not been widely explored. The literature on urban form and travel behavior has documented that the association between employment access and vehicle miles traveled (VMT) has one of the largest magnitudes among land use variables, with an elasticity of VMT with respect to employment access ranging from -0.05 to -0.31 (see, e.g., (Ewing and Cervero 2010) (Salon et al. 2012)). Since VMT is highly correlated with the level of greenhouse gas emissions, employment access variables have a potentially important policy role in sustainable urban development. Yet employment access variables in the research literature have not differentiated between whether drivers have access to dispersed jobs or jobs that are clustered in an employment sub-center. Clustering jobs in employment sub-centers might create agglomeration benefits that will affect household travel demand. We hypothesize that agglomeration benefits (in both production and consumption) within employment sub-centers are valuable to households, thus households that lack access to employment sub-centers may drive longer distances to access sub-centers, implying a weaker association between distance-based access to sub-centers and VMT. On the other hand, jobs that are centered will allow greater chaining of trips, implying that increased access to centered employment may reduce VMT more than increased access to non-centered employment. Overall, the net effect of access to employment sub-centers on VMT (relative to non-centered jobs) could be ambiguous. In this study, we separate centered and non-centered jobs in a traditional regression study of land use and VMT, to explore this question. We contribute to the literature by providing some of the first evidence about whether and how the association between employment access and VMT varies depending on whether households can access jobs in employment sub-centers or jobs that are not in sub-centers.

California metropolitan areas have highly sub-centered employment patterns, and the state has emphasized policy approaches that link land use to vehicle travel. We examine whether all jobs matter equally for travel behavior, and whether access to sub-centered jobs is or is not different (in terms of the association with household VMT) from access to jobs that are not in employment sub-centers. Our study area is the five-county Los Angeles Combined Statistical Area (CSA) – Los Angeles, Orange, Riverside, San Bernardino, and Ventura Counties. In this study, we use detailed data on employment from the National Employment Time Series (NETS), matched to geocoded firm locations, to identify employment sub-centers in the Los Angeles CSA. We then use the most recent travel diary survey for the region, the 2012 California Household Travel Survey (CHTS), to measure household VMT, again matching the household's residence to their geocoded location. In addition, we modify a standard land use – travel behavior regression to include, as explanatory variables, measures of access to jobs that are inside and outside of employment sub-centers. In order to examine questions of causality, we use access to historically significant places and transportation corridors to serve as instrumental variables for the access to jobs inside and outside of the contemporary employment centers. We use Tobit regressions to correct for the skewness of the distribution of household VMT because 25 percent of surveyed households have zero VMT on the survey day.

Our study shows that the access to jobs outside employment sub-centers has a larger association with household VMT than access to jobs within sub-centers. In addition, access to jobs inside smaller employment sub-centers has a larger association with household VMT than access to jobs in larger employment sub-centers. These findings show that the net effects between agglomeration benefits and trip chaining impacts are to make the elasticity of VMT with respect to access to jobs within sub-centers smaller than the elasticity of household VMT to access to jobs that are not within sub-centers. For local policy makers in the Los Angeles CSA, encouraging residential development in areas with higher access to non-centered jobs could be an effective method to reduce greenhouse gas (GHG) emissions. In addition, local governments might encourage job growth outside employment sub-centers, especially outside large employment sub-centers. However, this policy option would require local governments to sacrifice economic benefits from agglomeration to achieve sustainability goals. Section 2 of this paper reviews the relative literature, Section 3 introduces the data and methods of this study, Section 4 introduces the specification of the Tobit models, Section 5 shows and interprets the output of the models, Section 6 discusses the policy implication of the findings and Section 7 concludes.

## 2. Background and literature

### 2.1 Land use and driving

The literature on land use and driving has mostly been from cross-sectional regressions, with results summarized in the form of elasticities. For recent reviews, see, e.g., Boarnet (2011), Ewing and Cervero (2010), and Salon et al. (2012). These reviews all find that the elasticity of household VMT with respect to land use – travel behavior variables are in consistent ranges. The elasticity of household VMT with respect to residential density is typically in the range of -0.05 to -0.12, the elasticity of VMT with respect to access to employment is typically in the range of - 0.05 to -0.31 (see, e.g., Ewing and Cervero (2010); Salon et al. (2012)). The evolving consensus from these reviews is that access to employment (usually measured by gravity variables of the sort we use here) has a larger magnitude and hence is likely more important for VMT reduction than residential density.

The land use – travel behavior literature has also focused on questions of causality. Are relationships between policy and response variables (or between land use and travel variables) causal associations, or do those associations reflect omitted variables that lead travelers to live in land use settings that support their desired travel patterns? While a very few recent studies have used longitudinal designs to more clearly establish causality (Boarnet et al. 2017; Handy et al. 2006), with measured impacts that are often large, our study, like most, uses cross-sectional data due to lack of available panel data that can illuminate the question of sub-centered versus non-centered job access. For cross-sectional datasets, neighborhood-level socio-demographics were used to instrument density measures in regional-level studies (Boarnet and Sarmiento 1998), and geological characteristics were used to instrument landscape characteristics in national-level studies (Duranton and Turner 2017). Importantly, Duranton and Turner (2017) illuminate that there are two margins of endogeneity – potential sorting by households into residential locations based desired travel behaviors, and the possibility that land use measures (in our case, gravity variables of access to jobs) are correlated with the error term in a travel behavior regression. Several studies, including Duranton and Turner (2017), find that regressions that do not adjust for endogeneity (or sorting) of residential location give parameter estimates that are close to the results from instrumental variables (Cao et al. 2009). Based on Duranton and Turner (2017, Table 9), residential selection bias may be small and even zero, and the potentially larger magnitude of bias in a household VMT regression may be from endogeneity of land use characteristics. For that reason, we follow recent practice and use instruments to control for possible endogeneity of land use characteristics. Our instrumental variables approach is described more in Section 4.

## 2.2 Measuring access to employment

As discussed above, access to employment has a relatively large policy significance among land use variables. Employment access is usually measured by a gravity variable, which is the distance-weighted sum of jobs, for each geographic unit within a study area. The key element of the gravity access formulation is that, from any location, jobs are summed with a weight that is a function of the inverse of the distance between the household location and the jobs. Distant jobs matter less. This is the classic gravity formulation for access, so named because in early formulations, which interacted measures of employment and population and origins and destinations of trips, the mathematical formula was similar to the formula for gravitational potential energy (see, e.g., Haynes and Fotheringham (1984)).

Note that employment access proxies a broad range of trip destinations, not just work locations. Trip destinations are in very large part places where there are jobs, whether the trip is for work, education, entertainment, shopping, or services. Hence the employment access variable measures access to opportunities that should predict all travel, not just work-based (or commute) travel.

## 2.3 Polycentric urban structure

The spatial distribution of employment in cities is increasingly decentralized (Glaeser and Kahn 2001). In addition, dispersed employment has often concentrated around suburban employment centers. For instance, during the last two decades in the twentieth century, the Los

Angeles CSA has experienced rapid growth in suburban employment centers (Giuliano et al. 2007). In order to measure such increasingly decentralized urban spatial structure, employment sub-centers have been identified since the pioneering work of McDonald and McMillen (1990) and Giuliano and Small (1991). The early methods for identifying sub-centers divided a metropolitan area into geographic units (such as census tracts), and contiguous places with high employment densities were then aggregated, calling such an aggregation a sub-center if total employment and employment density were both above a threshold. In addition, some approaches use non-parametric methods to identify employment sub-centers (e.g. McMillen (2001)); Redfearn (2007)). This research uses the more straightforward parametric approach, described in the methods section, which conforms well to the approaches used by local planning agencies.

Employment sub-centers, with concentrated jobs, generally have higher productivity because of their agglomeration benefits (Puga 2010). Spatially concentrated employment can bring benefits in the form of shared infrastructure and labor pooling, better labor market matching, learning, and knowledge spillovers (Duranton and Puga 2004; Greenstone et al. 2010). In response to higher productivity benefits and agglomeration in consumption at employment sub-centers, household members might be willing to drive a longer distance to maintain access to employment sub-centers. In other words, VMT might be less elastic to access to jobs in employment sub-centers than access to jobs outside sub-centers, because as households are located farther from sub-centers they may still travel to sub-centers. Alternatively, the combination of opportunities in sub-centers may facilitate trip chaining that could imply a larger association between centered employment and household VMT than between non-centered employment and VMT. On net, we argue that theory gives reason to believe that jobs within and outside centers may influence household VMT differently, even if the net effect of which is larger (access to centered or non-centered jobs) cannot be determined *a priori*. The possibility that centered and non-centered employment accessibility have different magnitudes for household VMT has not yet been explored, and we address that gap in this paper.

Historical patterns of urban form and transportation infrastructure have been used as instrumental variables in the land use – transportation literature to identify causal relationships (e.g. (Baum-Snow 2007)). In the case of Los Angeles, the polycentric development pattern has been influenced by several historically important sites, including the location of the city center dating to the original Spanish settlement in the 1700s, the pre-World War II highway plans, and the street car systems that were important transportation arteries in the early 20$^{th}$ century (Brooks and Lutz 2014; Redfearn 2009). We use all of these historical locations to form instrumental variables, arguing that they measure variables that are likely correlated with the location of employment sub-centers (which have deep roots in any city's historical development) but that those historical variables are likely orthogonal to individual household driving decisions today. Thus, access to these sites can be used as instrumental variables for the current distribution of employment in the Los Angeles CSA.

### 2.4 State and regional policy contexts

The State of California has been actively seeking solutions to reduce GHG emissions. Climate change has been consistently observed in the atmosphere, ocean and cryosphere. Global-scale simulation models clearly show that human activity has contributed to global warming (Intergovernmental Panel on Climate Change 2013). The negative impacts of climate change include sea level rise, extreme weather and deteriorated ecosystems (Intergovernmental Panel on Climate Change 2013). As a pioneer in climate change mitigation in the United States, California passed *"The Global Warming Solutions Act"* (Assembly Bill 32, or AB 32) in 2006, requiring a reduction of state-level GHG emission to 1990 levels by 2020. In addition, *"The Sustainable Communities and Climate Protection Act"* (Senate Bill 375, or SB 375), passed into law in 2008, emphasizes using integrated land use and transportation planning as a policy tool to reduce GHG emissions. SB 375 requires metropolitan planning organizations (MPOs) to prepare "sustainable communities strategies" (SCS) as a part of its regional transportation plan to reduce regional level GHG emissions.

In the highly polycentric Los Angeles CSA, the local MPO (the Southern California Association of Governments, or SCAG) is using integrated land use and transportation planning as a policy tool to reduce VMT and GHG emission. In its most recent Regional Transportation Plan/ Sustainable Communities Strategy (2016-2040), SCAG is aiming to reduce GHG emission per capita by 8 percent by 2020, 18 percent by 2035 and 21 percent by 2040, compared to 2005 levels. In order to achieve these goals, SCAG plans to reduce VMT by promoting location efficient land use patterns and improving mass transit service (Southern California Association of Governments 2016). Los Angeles Metro, the largest mass transit provider in this region, is pursuing an ambitious expansion of its fixed-route rail transit network. Local voters recently passed a ballot initiative (Measure M) to increase the sales tax to support mass transit development. With the help of Measure M, the City of Los Angeles plans at least three major rail transit improvements finished by the time it hosts the 2028 Olympic Games (Nelson 2017). Amidst this aggressive planning, the link between land use and VMT in a polycentric region remains under-studied, in Los Angeles and elsewhere.

## 3. Methods and data

### 3.1 Employment sub-centers in the Los Angeles Combined Statistical Area

3.1.1 Identifying Employment Sub-centers

This study defines the Los Angeles Region as the Los Angeles Combined Statistical Area (CSA). The Los Angeles CSA consists of five counties: Los Angeles, Orange, Ventura, San Bernardino and Riverside. According to the most recent census in 2010, this area has 17,877,006 people, 48 percent of the total population in the State of California. The employment data come from the 2009 National Establishment Time-Series (NETS) Database, a proprietary dataset developed by Dun and Bradstreet (Walls and Associates 2008). The 2009 data are the most recent year available to the research team due to the nature of previous licensing agreements. We believe that the spatial pattern of employment sub-centers is likely to be stable over time, and that comparing 2012 travel (as described in Section 3.2) to 2009 patterns of employment is appropriate for the purposes of this study. The NETS database includes the geographic location (longitude and latitude) and the number of employees for each business establishment in the region.

We identified employment sub-centers in the Los Angeles CSA using the 2009 NETS database and the "95% - 10k" method introduced by Giuliano et al. (2015). We first divided the Los Angeles CSA into 34,527 hexagons, such that each hexagon has an area of 640 acres or one square mile. The employment centers contain hexagons with employment density larger than the $95^{th}$ percentile of the entire region, or 1,115 jobs per square mile. Contiguous hexagons with employment densities above the region's $95^{th}$ percentile are grouped together into candidate sub-centers. When those contiguous collections of high density sub-centers have a total of at least 10,000 jobs, the location is identified as an employment sub-center. Using this "95% - 10k" method we identified 46 employment centers in the Los Angeles CSA (FIGURE 1), which will be discussed in detail in the next section.

### 3.1.2 Distribution of employment sub-centers

The 46 employment sub-centers have 3,331,205 total jobs, 39.8 percent of the total 8,366,369 jobs in the Los Angeles CSA. Compared to a similar study using the 1980 Census Journey-to-Work data by Giuliano and Small (1991), employment is more concentrated in the employment sub-centers in 2009 than in 1980. In the 1980 study, Giuliano and Small (1991) identified 32 employment sub-centers in the same region, containing 32.1% of the region's total employment.

The largest employment sub-center in the Los Angeles CSA is a corridor extending from downtown Los Angeles westbound to Santa Monica through the Wilshire Corridor. This sub-center contains 1,107,139 total jobs, 33.2 percent of the total jobs located inside all employment sub-centers and 13.2 percent of the total jobs in the Los Angeles CSA. The second largest employment sub-center is in the heart of Orange County, extending from Anaheim to Irvine through Santa Ana. This employment sub-center has 605,284 jobs, 18.2 percent of the total jobs located inside the employment sub-centers and 7.2 percent of the total jobs in the Los Angeles CSA. The other 44 employment sub-centers have employment ranging from 162,332 for the third largest sub-center, to 10,081 for the 46$^{th}$ largest sub-center. The third to 44$^{th}$ largest employment centers have an average of 36,791 jobs per sub-center. The third, fourth and fifth largest employment sub-centers are located in the South Bay, San Fernando Valley and Los Angeles International Airport areas, respectively.

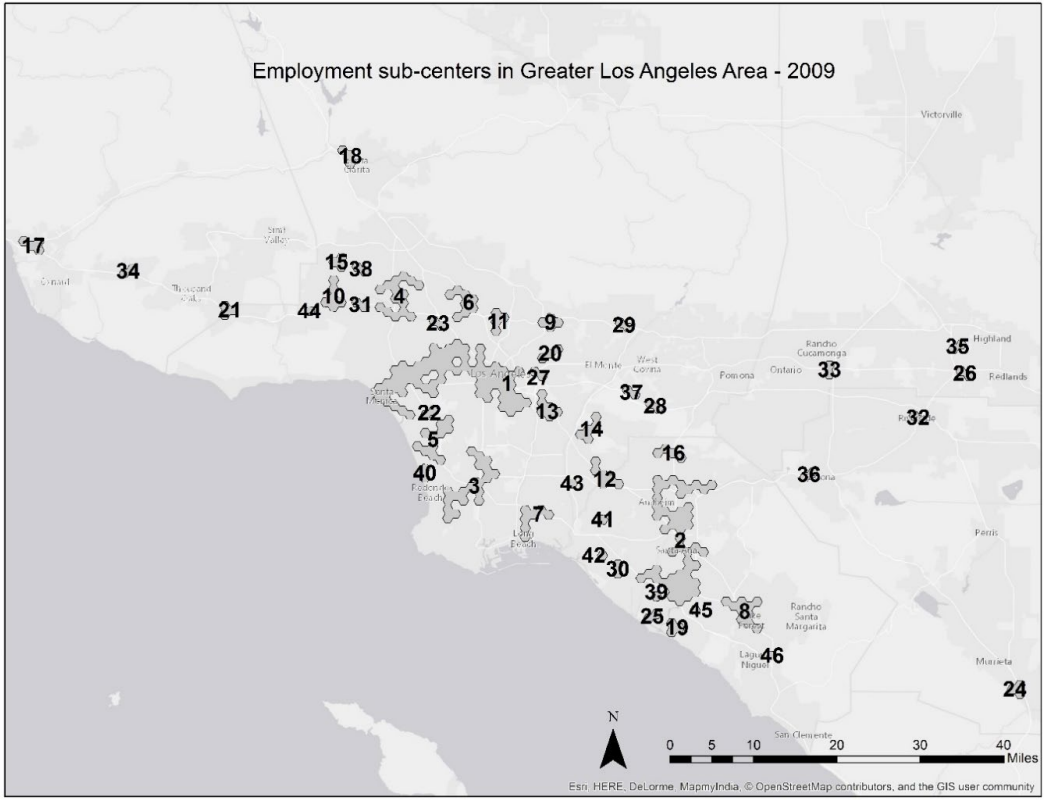

**FIGURE 1 Employment sub-centers in Los Angeles CSA**
Note: numbers 1 through 46 indicate the rank of the size of employment for each sub-center

FIGURE 1 also shows that 37 of the 46 employment sub-centers in the Los Angeles CSA are located in the two most urbanized counties – Los Angeles and Orange. There are three employment sub-centers located inside each of the other three suburban or exurban counties: Riverside, San Bernardino and Ventura.

### 3.2 Household VMT

The household VMT data are from the 2012 California Household Travel Survey (CHTS), the most recent statewide travel survey conducted by the California Department of Transportation. The CHTS used an activity diary that captured information about trips. Each household member was asked to estimate trip length for every trip during her or his diary day.

We used those data to calculate VMT for households, aggregating trip length for all trips made in household vehicles, taking care not to double-count VMT for trips in vehicles with multiple household members. Trips in vehicles not owned by the household are also counted in the sum of household VMT. We considered households with a VMT higher than 200 on the survey day to be outliers and removed them. In the Los Angeles CSA, 14,877 households were surveyed in the CHTS. The average daily household VMT in the sample is 35.6 and the standard deviation is 40.9. However, more than 25 percent of the households have zero VMT in the survey day. Such a distribution indicates that our sample is left-censored and using an ordinary least squares (OLS) model is not appropriate. We use Tobit regression in our analysis, which will be discussed in detail in Section 4.

### 3.3 Measuring access to jobs inside and outside employment sub-centers

We created a gravity-type job accessibility index to measure the impact of employment sub-centers on individual household VMT. Mathematically, the value of the job accessibility index is dependent on two factors: the number of jobs in the region and the distance from the jobs to a resident's hexagon. The job accessibility index is defined as the number of jobs damped by the distance between the jobs and the residence. We use hexagons as the unit of analysis to create the accessibility variable, as shown below:

$$emp\_acc_i = \left(\sum_{j \neq i} \frac{E_j}{D_{ij}^2} + E_i\right) \cdot \frac{1}{10,000} \quad (1),$$

where *emp_acc$_i$* refers to the job accessibility index of hexagon *i*, $E_j$ refers to the number of jobs inside hexagon *j*, $D_{ij}$ refers to distance (in miles) between the centroids of hexagon *i* and hexagon *j*, and $E_i$ refers to number of jobs inside hexagon *i*. In other words, the accessibility index of hexagon *i* equals the number of jobs within that hexagon, plus the sum of jobs of the other hexagons in the region with quadratic distance damping.

As discussed previously, the Los Angeles CSA is highly polycentric with 46 employment sub-centers. The jobs in these employment sub-centers constitute 39.8 percent of the total employment in the region. As discussed in Section 2.3, access to employment located inside and outside employment sub-centers might impact household VMT differently. In order to test this hypothesis, we break the job accessibility index into multiple indices, measuring accessibility to jobs inside and outside employment sub-centers. In addition to the accessibility to all employment in the region, we first create two accessibility indices for accessibility to jobs located inside and outside employment sub-centers. Each index is a gravity summation of access to employment from each hexagon *i*, with the former (access to jobs in sub-centers) only summed over jobs that are in the 46 sub-centers, while the latter is summed over all other hexagons (those not in sub-centers). The jobs in the 46 employment sub-centers are highly concentrated in the two largest sub-centers. The largest and second-largest employment sub-centers contain 33.2 percent and 18.2 percent of the total jobs located inside employment sub-centers, respectively. The size of the employment sub-centers might impact the level of the agglomeration economies and possibly the nature of the land use – travel interaction (e.g. from trip chaining possibilities) from employment access to VMT. Thus, we further break the accessibility variables into three different indices: accessibility to jobs within only the largest employment sub-center, accessibility to jobs within the second-largest employment sub-center, and accessibility to jobs within the third-to-46[th] largest employment sub-centers.

The descriptive statistics for these indices are shown as TABLE 1 below. The areas with the highest access to jobs in the sub-centers are located within the employment sub-centers (FIGURE 2-a), while the areas with the highest access to jobs outside the sub-centers are located between major sub-centers (FIGURE 2-b).

TABLE 1: Descriptive Statistics of Employment Accessibility Indices

| Accessibility variables | N | Mean | Std. Dev. | Min | Max |
|---|---|---|---|---|---|
| to all jobs | 14,877 | 6.264 | 4.648 | 0.023 | 32.84 |
| to jobs in the sub-centers | 14,877 | 3.151 | 4.008 | 0.007 | 29.85 |
| to jobs outside sub-centers | 14,877 | 3.113 | 1.478 | 0.016 | 6.652 |
| to jobs in the largest sub-center | 14,877 | 1.526 | 3.657 | 0.002 | 28.745 |
| to jobs in the second-largest sub-center | 14,877 | 0.377 | 1.192 | 0.001 | 17.067 |
| to jobs in the third-to-46th largest sub-centers | 14,877 | 1.247 | 1.395 | 0.003 | 7.922 |

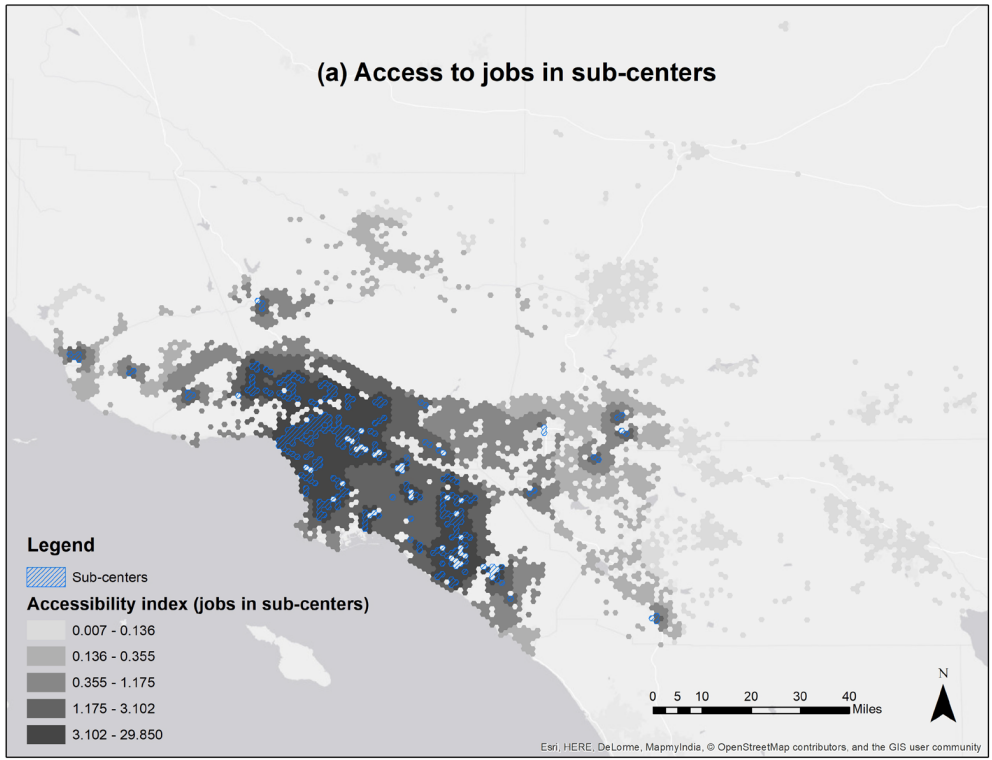
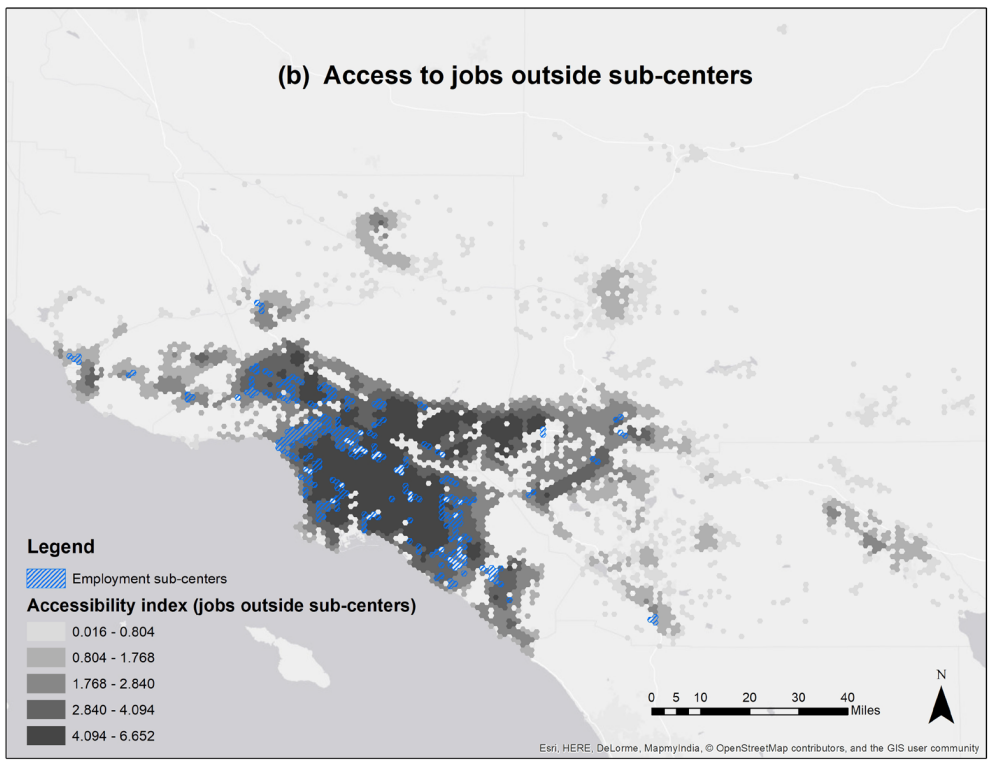

**FIGURE 2** – Access to centered jobs (a) and non-centered jobs (b)

## 4. Model specification

We estimated three sets of regression models, following the equation below:

$$VMT_i = \beta_0 + \mathbf{acc}_i' \cdot \boldsymbol{\beta_1} + \mathbf{X}_i' \cdot \boldsymbol{\beta_2} + \varepsilon_i \quad (2),$$

where, for household $i$, $VMT_i$ indicates the household-level daily VMT, column vector $\mathbf{acc}_i$ indicates the job accessibility variables, and the column vector $\mathbf{X}_i$ indicates control variables. The coefficients are the scalar $\beta_0$ and the column vectors $\boldsymbol{\beta_1}$ and $\boldsymbol{\beta_2}$. Specifically, the vector $\mathbf{X}_i$ includes: number of vehicles in the household, household income (dummy variables in ten categories) household size and population density of the census tract where the household lives. Each observation in the regression is a household, and the access variables are all generated relative to the household's location (their hexagon of residence) and so measure employment access from that household's location. The descriptive statistics of VMT and other socio-demographic variables are in TABLE 2 below. TABLE 2 shows that the average household VMT is 35.55, and 70 percent of the respondents reside in the two most urbanized counties: Los Angeles and Orange.

**TABLE 2: Descriptive Statistics of Dependent and Independent Variables**

| Variable | N | Mean | Std. Dev. | Min | Max |
|---|---|---|---|---|---|
| household VMT | 14,877 | 35.55 | 40.9 | 0 | 199.99 |
| number of vehicles in household | 14,877 | 1.85 | 1 | 0 | 8 |
| household income category dummy variables | | | | | |
| *less than $ 10,000* | | (reference term) | | | |
| *$10,000 to $24,999* | 13,475 | 0.13 | 0.33 | 0 | 1 |
| *$25,000 to $34,999* | 13,475 | 0.09 | 0.28 | 0 | 1 |
| *$35,000 to $49,999* | 13,475 | 0.12 | 0.32 | 0 | 1 |
| *$50,000 to $74,999* | 13,475 | 0.17 | 0.38 | 0 | 1 |
| *$75,000 to $99,999* | 13,475 | 0.15 | 0.36 | 0 | 1 |
| *$100,000 to $149,999* | 13,475 | 0.17 | 0.37 | 0 | 1 |
| *$150,000 to $199,999* | 13,475 | 0.07 | 0.26 | 0 | 1 |
| *$200,000 to $249,999* | 13,475 | 0.03 | 0.17 | 0 | 1 |
| *$250,000 or more* | 13,475 | 0.03 | 0.18 | 0 | 1 |
| household size | 14,877 | 2.66 | 1.44 | 1 | 8 |
| population density of the census tract where the household lives (1,000 per square mile) | 14,877 | 8.52 | 8.07 | 0.001 | 94.49 |
| flag for Los Angeles and Orange Counties | 14,877 | 0.7 | 0.46 | 0 | 1 |

In order to control for the possibility that the employment access variables are endogenous to household VMT, possibly through unobservable factors associated with both job accessibility and driving, we instrumented for the access variables. We are not able to control for the endogeneity from residential sorting, which, as argued by Duranton and Turner (2017), does not primarily contribute to the urban form-driving relationship. We selected four gravity accessibility variables to serve as instrumental variables. Each variable represents distant historical characteristics of the region which are correlated with present-day employment sub-centers but which, we assume due to the long-time lag, are not correlated with the error term in the VMT regression. The instruments are: accessibility to Pacific Electric Railway (Red Car) stations in 1926, accessibility to Los Angeles Railway (Yellow Car) lines in 1906, accessibility to planned highways in 1939 (as planned, not as built), and inverse distance to the Los Angeles Plaza built in the late 1700s by Spanish settlers. As discussed in Section 2.3, historical sites and transportation corridors are important driving forces for the current polycentric urban structure of the Los Angeles CSA (Brooks and Lutz 2014; Redfearn 2009). Both the Red Car and the Yellow Car were streetcar systems, while the former is a regional one and the latter mainly served the central city. The Red Car had fixed stations and the Yellow Car did not (the Yellow Car stopped at locations on streets based on passenger requests). Both systems started in the 1890s, peaked in the 1910s and experienced declines until complete closure in the 1960s (Brooks and Lutz 2014;

Redfearn 2009). The Red Car map is digitized from a map (Electric Railway Historical Association of Southern California 2017) showing the 1926 system (Walker 2006), and the Yellow Car digital map (sharemap.org 2017) shows its system in 1906 (Travel and Hotel Bureau 1906). The planned highway system as of 1939 is from the City of Los Angeles Transportation Engineering Board (1939). The Los Angeles Plaza, the geographical center of the early Spanish settlement in the late 1700s, is georeferenced by Google Map.

All these variables are calculated using hexagons as a unit, the accessibility variables are calculated following the equation below:

$$iv\_acc_i = \sum_{j \neq i} \frac{A_j}{D_{ij}^2} + A_i \quad (3),$$

where $A_i$ and $A_j$ are dummy variables equal 1 if the hexagon has Red Car stations, Yellow Car lines or 1939 planned highways passing through it, and 0 otherwise. $D_{ij}$ refers to distance (in miles) between centroids of hexagon $i$ and hexagon $j$. This equation follows the same logic of Equation (2). There are four households sharing the same hexagon with Los Angeles Plaza, their inverse hexagon-to-hexagon distances are set to be 1, since the values for the rest 14,876 households are no larger than 0.931. The maps showing the hexagons used to calculate the instrumental variables are shown as FIGURE 3 below.

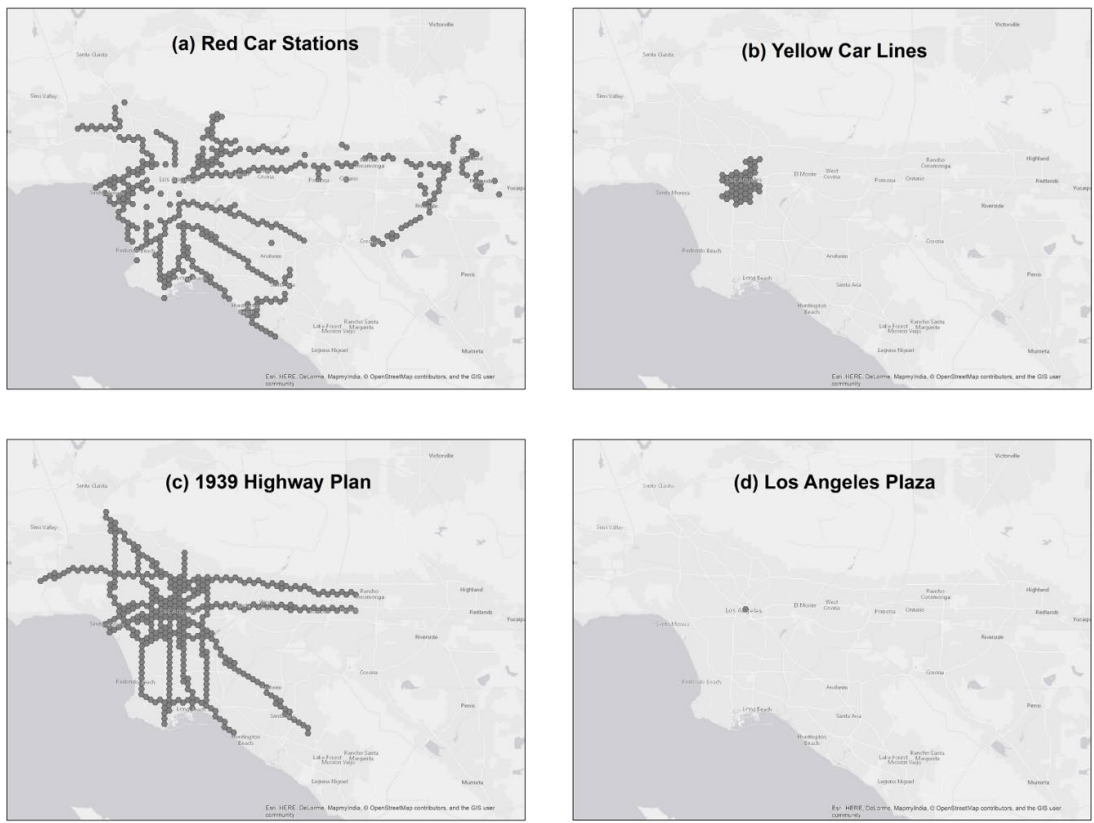

**FIGURE 3 – Hexagons used to calculate instrument variables**
Notes: Map of the Red Car stations is from the Electric Railway Historical Association of Southern California (2017), Map of the Yellow Car lines is from sharemap.org (2017), Map of the 1939 Highway Plan is from City of Los Angeles Transportation Engineering Board (1939), and the location of the Los Angeles Plaza is georeferenced by Google Map.

We estimated three sets of models with different access to jobs indices. The first set focuses only on accessibility to all jobs. In the second set we broke the single index into two variables to measure accessibility to jobs inside and outside sub-centers. In the third set we further broke the accessibility variable into measures of access to jobs within the largest, second-largest, and the 3$^{rd}$-46$^{th}$ largest sub-centers. In each group of regressions, we estimated a model for all households and a model only for households in the two most urbanized counties (Los Angeles and Orange). As discussed in Section 3.2, the distribution of household-level VMT is left-censored, more than 25 percent of the households in our sample have zero VMT on the survey day. Such a distribution violates the basic assumption of the ordinary least square (OLS) models, and Two-stage Least Square (2SLS) models will give biased estimation results. Instead, we used basic Tobit models and Two-stage Tobit models with instrumental variables. Tobit models are designed to estimate regressions with censored dependent variables using a Maximum Likelihood Estimation method.

To provide a unit-free measurement of the magnitude of impact of accessibility on household-level VMT, we estimated elasticities based on the regression model outputs. An elasticity value of *x* can be interpreted as the percentage change in the dependent variable (VMT) associated with a one-percent change in the independent variable of interest. Since the Tobit model is not linear, we estimate the elasticity of household VMT with respect to accessibility following the method of Boarnet et al. (2011) shown in the formula below:

$$e = \frac{1}{n} \sum_i me_i * \frac{accessibility_i}{VMT_i} \quad (4),$$

where *e* refers to elasticity, $me_i$ refers to the marginal effect for household *i*, which equals the predicted probability of household *i* having a positive VMT multiplied by the estimated coefficient of the accessibility variable, $accessibility_i$ and $VMT_i$ are the values for the *i*'th household, and *n* refers to number of observations.

## 5. Results

Outputs of the models for household VMT regressed on access to (a) all jobs, (b) jobs inside and outside employment sub-centers, and (c) jobs inside the largest sub-center, the second-largest sub-center, the third-to-46$^{th}$ largest sub-center and outside sub-centers, are shown in TABLE 3, TABLE 4 and TABLE 5, respectively. In each table, the control variables shown in Equation (2) are included, but the coefficients for control variables are not reported for brevity. Full results are available from the authors upon request. In these tables, we present results of both the Tobit models and the Two-stage Tobit models. For each model specification, we present one for the full sample and the results estimated only for households in the two most urbanized counties of Los Angeles and Orange. For the third set of models, we are not able to build Two-stage Tobit models because of a lack of enough instrumental variables to perform over-identification tests.

We tested the validity of our instruments with over-identification tests (as shown in TABLE 3 and TABLE 4). The Stata software package does not offer an over-identification test for Two-stage Tobit models. Instead, for each of the four Two-stage Tobit models, we use the result of the over-identification test of its corresponding 2SLS model as a reference. The 2SLS counterparts of all of the four Two-stage Tobit models have passed the over-identification tests; in no case do the 2SLS counterparts of the Two-stage Tobit models reported here reject the null of valid instruments. In addition, there is a possibility that the historical characteristics that we use as instruments might be associated with patterns of transportation infrastructure in the present day which could create correlation between the instrument and the error term in the VMT regression. To examine that, we added a variable that measured the number of highways that pass through the census tract of residence to the VMT regression. Including that variable did not change the sign or significance on any of the sub-center variables, providing additional assurance that there is not remaining correlation between present-day highways and the error in the VMT regression, which might be a path through which the instruments may be correlated with the error in the VMT regression. (These regression results are available on request from the authors.)

Admittedly, our instruments can control for the endogeneity of sub-center locations, but not for residential selection. We note that Duranton and Turner (2017), in a national study of VMT in relation to density, estimate an upper bound for residential selection bias of 17 percent, which is consistent with the idea that VMT regressions that control for household sociodemographics likely do not suffer from residential selection bias (see, e.g., Brownstone (2008) for this argument). To further examine the question of residential selection, we examine whether there are omitted variables that may bias the results, following the reasoning of Brownstone (2008) and Duranton and Turner (2017) that the absence of such omitted variable bias provides some assurance that residential selection is not an issue.

We implemented a simple test on omitted variable bias suggested by Oster (2017) for Model 1 (full sample, access to all jobs in TABLE 3) and Model 5 (full sample, access to centered and non-centered jobs in TABLE 4). Specifically, we used the "restricted estimator" assumption in Oster (2017) and compared the coefficients on the job access variables with and without household characteristics. Both Oster-corrected and uncorrected coefficients had the same sign, and the patterns of estimated coefficients still show that access to non-centered jobs has a larger effect than access to non-centered jobs. (Results available on request from the authors.) This provides reason to believe that even if there is residential selection bias, it is unlikely to affect our primary result, which is about comparisons of the magnitude of the resulting elasticities of VMT with respect to centered and non-centered jobs.

TABLE 3 shows that accessibility to all jobs is negatively associated with household VMT. In TABLE 3, Model 1 and Model 3 are baseline Tobit models, and Model 2 and Model 4 are Two-stage Tobit models with two instrumental variables (access to Yellow Car lines and inverse distance to Los Angeles Plaza). Model 2 shows that the elasticity of VMT with respect to access to all jobs is -0.127. Model 4 shows that for households residing in the two most urbanized counties, the elasticity is larger in magnitude: -0.156. Compared with Models 2 and 4, the Tobit models (Models 1 and 3) have larger absolute values of elasticities, showing that Tobit models overestimate the VMT effects of access to all jobs.

TABLE 3 – Household VMT and access to all jobs

|  | Model 1 Tobit | Model 2 IV Tobit | Model 3 Tobit | Model 4 IV Tobit |
|---|---|---|---|---|
| access to all jobs | -0.815*** | -0.533*** | -0.854*** | -0.495** |
|  | (0.116) | (0.199) | (0.137) | (0.237) |
| Elasticity | -0.195 | -0.127 | -0.270 | -0.156 |
| p-value for joint significance | <0.001 | <0.001 | <0.001 | <0.001 |
| p-value for over-identification test of corresponding 2SLS model | N/A | 0.192 | N/A | 0.354 |
| sample | all households | all households | Los Angeles and Orange Counties | Los Angeles and Orange Counties |
| # of observations | 13,745 | 13,745 | 9,361 | 9,361 |

Notes: Dependent variables for all models are household VMT. Instrumental variables for Model 2 and Model 4 are access to Yellow Car lines and inverse distance to Los Angeles Plaza. Control variables are numbers of vehicles in the household, household income categories, household size and population density of the census tract where the household lives. Robust standard errors in parentheses. All models passed likelihood ratio tests for joint significance. The F-statistics of the first-stage regressions for Model 2 and Model 4 are both larger than 10. *, **, *** indicate significance at 10%, 5% and 1% levels, respectively.

TABLE 4 shows that access to jobs outside sub-centers has a larger effect on household VMT than access to jobs inside employment sub-centers. Models 5 and 7 are baseline Tobit Models, and Models 6 and 8 are Two-stage Tobit models, with three instrumental variables: access to 1926 Red Car stations, access to 1906 Yellow Car lines and access to the 1939 planned highways. Model 6 shows that, for all households in the region, access to jobs inside and outside employment sub-centers are both negatively associated with household VMT. The elasticity of VMT with respect to access to jobs inside sub-centers is -0.084, while that of access to jobs outside sub-centers is -0.149. Model 8 for the households in the two most urbanized counties also shows that access to jobs inside employment sub-centers has a smaller effect on household

VMT than access to jobs outside sub-centers, with elasticity of -0.104 and -0.364, respectively. Compared to Models 6 and 8, the Tobit models (Models 5 and 7) tend to overestimate the effect of centered jobs and underestimate the effect of non-centered jobs. Also, for both measures of access to jobs, the magnitude of their effects on household VMT is larger for the households in Los Angeles and Orange counties. This pattern is similar to what we have found in TABLE 3 for models on access to all jobs.

TABLE 4 – Household VMT and access to jobs inside and outside employment sub-centers

|  | Model 5 Tobit | Model 6 IV Tobit | Model 7 Tobit | Model 8 IV Tobit |
|---|---|---|---|---|
| access to jobs inside employment sub-centers | -0.722*** (0.127) | -0.700*** (0.256) | -0.784*** (0.136) | -0.604** (0.255) |
| Elasticity | -0.087 | -0.084 | -0.135 | -0.104 |
| access to jobs outside employment sub-centers | -1.258*** (0.303) | -1.301** (0.641) | -1.863*** (0.424) | -2.525** (1.076) |
| Elasticity | -0.149 | -0.155 | -0.269 | -0.364 |
| p-value for joint significance | <0.001 | <0.001 | <0.001 | <0.001 |
| p-value for over-identification test of corresponding 2SLS model | N/A | 0.160 | N/A | 0.234 |
| Sample | all households | all households | Los Angeles and Orange Counties | Los Angeles and Orange Counties |
| # of observations | 13,745 | 13,745 | 9,361 | 9,361 |

Notes: Dependent variables for all models are household VMT. Instrument variables for Model 6 and Model 8 are access to Red Car stations, access to Yellow Car lines and access to 1939 planned highway. Control variables are numbers of vehicles in the household, household income categories, household size and population density of the census tract where the household lives. Robust standard errors in parentheses. All models passed likelihood ratio tests for joint significance. The F-statistics of the first-stage regressions for Model 6 and Model 8 are both larger than 10. *, **, *** indicate significant at 10%, 5% and 1% levels, respectively.

TABLE 5 shows that access to employment in smaller sub-centers has stronger associations with household VMT than access to employment in larger sub-centers. In Model 9, the magnitudes of elasticity on household VMT for access to jobs in the largest, the second largest and the 3$^{rd}$-46$^{th}$ largest sub-centers are -0.037, -0.013 and -0.058, respectively. Similarly, in Model 10 for households in the two most urbanized counties only, the magnitudes of elasticity on household VMT for access to jobs in the largest, the second largest and the third to 46[th] largest sub-centers are -0.059, -0.023 and -0.095, respectively. TABLE 5 also shows that for each variable, the magnitudes of elasticity in Model 10 (only for Los Angeles and Orange Counties) is larger than those in Model 9 (for the whole sample). Because of a lack of enough instrumental variables to perform over-identification tests, we cannot build Two-stage Tobit models to establish causal relationships for the models in TABLE 5. Nevertheless, based on the fact that the significance does not change when moving from Tobit to Two-stage Tobit models for Models 1-8, it is very possible that such a pattern will persist in Model 9 and Model 10.

**TABLE 5 – Household VMT and access to jobs inside the largest sub-center, the second largest sub-center, the 3rd-46th largest sub-centers and outside sub-centers**

|  | Model 9 Tobit | Model 10 Tobit |
|---|---|---|
| access to jobs inside the largest employment sub-center | -0.640*** (0.136) | -0.697*** (0.141) |
| elasticity | -0.037 | -0.059 |
| access to jobs inside the second-largest employment sub-center | -0.926*** (0.294) | -1.128*** (0.299) |
| elasticity | -0.013 | -0.023 |
| access to jobs inside the third-to-46th largest employment sub-centers | -1.196*** (0.294) | -1.418*** (0.310) |
| elasticity | -0.058 | -0.095 |
| access to jobs outside employment sub-centers | -1.041*** (0.324) | -1.640*** (0.430) |
| elasticity | -0.124 | -0.237 |
| p-value for joint significance | <0.001 | <0.001 |
| sample | all households | Los Angeles and Orange Counties |
| # of observations | 13,745 | 9,361 |

Notes: Dependent variables for both models are household VMT. Control variables are numbers of vehicles in the household, household income categories, household size and population density of the census tract where the household lives. Robust standard errors in parentheses. All models passed likelihood ratio tests for joint significance. *, **, *** indicate significant at 10%, 5% and 1% levels, respectively.

To summarize, access to employment is negatively associated with household VMT in the Los Angeles CSA. Where households have better access to employment, they drive less, *ceteris paribus*. The elasticity of the effect is in the range from -0.04 to -0.36, varying by samples and the specific measure of access. The magnitude of the effect of access to jobs outside employment sub-centers on household VMT is larger than that for jobs inside employment sub-centers. The magnitude of the elasticity of household VMT with respect to access to jobs in smaller sub-centers is larger than that in larger sub-centers. In addition, for each measure of accessibility, the magnitude of association with household VMT is larger for households in the two most urbanized counties (Los Angeles and Orange).

## 6. Policy implications

The outputs of the models presented in Section 5 show that access to jobs outside employment sub-centers has the largest association with household VMT. Household that have better access to non-centered jobs drive less, and while access to centered jobs also reduces driving, access to non-centered jobs in the Los Angeles CSA gives a larger elasticity of VMT reduction. Although numerical policy simulation is beyond the scope of this paper, this section provides some qualitative discussion of the policy implications of the results of this study.

We start with the observation that the result that access to non-centered jobs does more to reduce VMT will seem counterintuitive to many planners. Ideas of compact cities and sustainable urban development have often focused on creating mixed-use centers – admittedly not the same as employment sub-centers, but potentially consistent with adding residential development to sub-centers. For instance, SCAG's Regional Transportation Plan/ Sustainable Community Strategy (2016-2040, Page 8) aims to accommodate 46 percent of the region's future household growth in "High Quality Transit Areas", comprising only three percent of the total land area of the region (Southern California Association of Governments 2016). Additionally, firms locate in sub-centers likely in part to benefit from agglomeration economies, and so decentralizing sub-centers would at best have to balance the costs and benefits of congestion versus agglomeration and at worst may make little sense. A full analysis of these issues is beyond the scope of this paper. Instead, let us start with an examination of what locations have the best access to non-centered jobs in the Los Angeles CSA.

FIGURE 4 shows the top quintile in access to non-centered jobs of the 3,392 hexagons where the households in our sample reside. As the map indicates, most of those areas are outside employment sub-centers, but usually within the two most urbanized counties of Los Angeles and Orange. The locations in the top quintile of access to non-centered jobs are locations that are often prime candidates for infill development, near major sub-centers but in locations such as the South Bay communities, north Orange County, and the San Gabriel Valley that are first generation suburbs. These are locations that are accessible, interestingly, to both non-centered and centered jobs. The policy recommendation flowing from this analysis is not to develop on the fringe, but (from FIGURE 4) to fill in the gaps between sub-centers, possibly with residential development that is well connected to the nearby employment sub-centers. Such a strategy is, in fact, consistent with SCAG's approach to focus on infill in the urbanized core, but with a modification that the densest locations (e.g. downtown Los Angeles or the Westside's Wilshire corridor) may not be the best opportunities for VMT-reducing infill development. Rather, the results (again, see FIGURE 4) suggest a focus on locations that are infill opportunities that are near, but not within, sub-centers. We note that this is a qualitative interpretation, and we call for future land use simulation research to examine how the evidence in this paper informs the relationship between future growth patterns and VMT reduction.

Note that some of the areas in the top quintile of access to non-centered jobs are close to rail transit, bus rapid transit (BRT) or commuter rail (Metrolink) stations. Policies encouraging residential development in such areas, especially around rail transit, BRT and commuter rail stations, could be helpful in reducing VMT and hence GHG emissions in this region. A recent study has proved that people residing closer to light rail stations in the city of Los Angeles have lower GHG emissions from private vehicles than those living farther away from those stations (Boarnet et al. 2017). This provides reason to believe that transit access can be used to leverage development in the locations that are accessibility to non-centered jobs in FIGURE 4, connecting those locations to nearby employment sub-centers through infrastructure that does not require driving. This also conforms to SCAG's aforementioned strategies to prioritize infill development in areas with good access to mass transit.

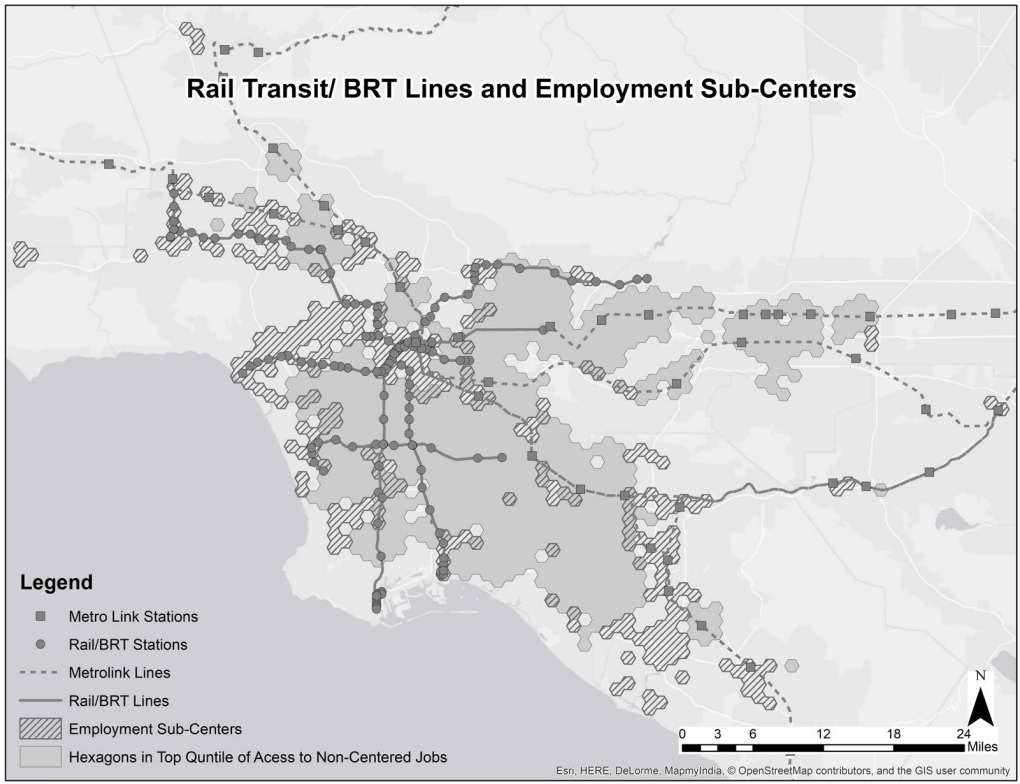

**FIGURE 4 – Rail transit, BRT Lines and Employment Sub-centers**

Another potential policy to increase access to non-centered jobs is to encourage business development outside employment sub-centers or inside smaller sub-centers. This implication is related to the finding from Cervero and Duncan (2006) that job-housing imbalances are related to higher work-related VMT, possibly by separating job locations from residence locations. However, we believe that this policy direction is likely less fruitful than the first one. Such a policy might change the spatial distribution of employment and hence the distribution of the sub-centers. The net effect on household VMT for this policy is unclear without the help of numerical simulation models. In addition, policies encouraging a more even distribution of employment might reduce the agglomeration benefits of large employment sub-centers, forcing local policy makers to trade off economic development and sustainability objectives.

# 7. Conclusions

This study has investigated the relationship between urban polycentricity and household VMT, using the Los Angeles CSA as an example. We have identified 46 employment sub-centers in this region, and estimated the effect of access to jobs inside and outside these 46 sub-centers on household VMT. To examine a causal relationship and control for the possibility that the error term in a household VMT regression is correlated with access to centered and non-centered jobs, we use access to historically important places and transportation infrastructure as instrumental variables for our gravity measures of job access. Our Two-stage Tobit models found that access to jobs inside and outside employment sub-centers are both negatively associated with household VMT, while the latter effect is larger in magnitude than the former. We have also found that access to employment in smaller sub-centers has larger associations with household VMT than access to employment in larger sub-centers. For any specific measurement of access to jobs, the magnitude of its effect on household VMT is larger for households living in the two most urbanized counties: Los Angeles and Orange. This study adds to the existing literature on urban form and travel by examining the impact of polycentric urban structure on household VMT. To our knowledge, this study is among the first to study the effects of access to jobs within and outside employment sub-centers on urban passenger travel.

The findings of this study reinforce previous research by suggesting the potential for integrated land use and transportation planning as a policy tool to reduce GHG emission from the transportation sector. On average, doubling the access to non-centered jobs could reduce household VMT by 16 percent in Los Angeles CSA. For instance, for the areas in the top quintile of access to non-centered jobs (shown in FIGURE 4), the indices of access to non-centered jobs are from 4.10 to 6.65. For areas in the inland Riverside County adjacent to centers numbered 26, 32 and 35 (the numbered IDs are shown in FIGURE 1), the indices fall to half of those values (2.05 to 3.33). Holding other factors constant, moving households from these inland areas to the areas in the top quintile of access to non-centered jobs could generate approximately a 16-percent VMT reduction. For local policy makers in the Los Angeles CSA, encouraging residential development in areas with high accessibility to non-centered jobs, especially in such areas with good access to rail transit, BRT and commuter rail, could be effective in reducing VMT and hence GHG emissions. This finding confirms the policy direction of the region's local MPO in its most recent regional transportation plan (Southern California Association of Governments 2016). In addition, local governments could also think about encouraging business development outside large employment sub-centers to improve the region's job-housing balance (Cervero and Duncan 2006). However, that might reduce the agglomeration benefits of large employment sub-centers, and hence we do not suggest that as a policy direction. Admittedly, numeric simulations are needed to quantify the transportation effects of these policy directions, since these policies will change the spatial distribution of both population and jobs.

      We acknowledge that our instruments can control for the endogeneity of sub-centers, but not for residential selection. In national research, Duranton and Turner (2017) found that residential selection biases the coefficient on a density variable (in a VMT regression) by at most 17 percent. We are not able to implement the correction for residential selection used in Duranton and Turner (2017), because that requires historical data on the key land use variable, and we can only construct measures of sub-center location for one time period due to the data licensing agreement that we used for our employment data. We note that, from Duranton and Turner (2017) and other authors (e.g. Cao et al., 2009), the magnitude of residential selection bias is typically small in travel behavior regressions. Our simple test following Oster (2017)

indicated that if there is residential selection bias, it is unlikely to affect our result with respect to the different effects of VMT of access to centered and non-centered jobs.

However, our Two-stage Tobit estimates might still overestimate the effect of job access on VMT because we were not able to deal definitively with the endogeneity from household sorting into residential locations. Also, the effects of accessibility on driving might be confounded with amenities such as schools and parks. Nevertheless, our elasticity estimates are within the range of other estimates in the literature, but with additional ability to distinguish between different kinds of job access and hence provide more fidelity in policy suggestions. While we caution that we cannot establish that these associations as causal, we also note that the large literature on land use and travel has much evidence that correcting either for sorting or for the endogeneity of land use does not change coefficient magnitudes much. We draw policy inferences as if the association is causal, with the caution that while there is a growing preponderance of the evidence in the literature suggesting such an inference is reasonable, causal identification in this realm remains an open issue.

Future research should focus on expanding this study to the national level, with national-level employment data, to examine these relationships in other metropolitan areas with different industrial structures and land use patterns. A national-level, trip-based dataset could help to examine the effects of employment sub-centers with different industrial patterns on VMT with different trip purposes. Additionally, leveraging longitudinal data (still rare in travel behavior) can help examine changes in driving behavior of households who recently moved, which can further help illuminate the causal relationships that are in play.

## Acknowledgements

This study was funded by California Department of Transportation through the METRANS Transportation Center in task order 005-A01. The authors thank the Editor and the anonymous referees for their helpful comments.